\documentclass[%
 aip,
 amsmath,amssymb,prl,
 reprint,%
]{revtex4-2}

\usepackage{graphicx}
\usepackage{dcolumn}
\usepackage{bm}

\usepackage[utf8]{inputenc}
\usepackage[T1]{fontenc}
\usepackage{mathptmx}
\usepackage{mathrsfs}
\usepackage{etoolbox}
\usepackage{xcolor,soul}
\sethlcolor{yellow}
\usepackage{cancel}

\usepackage[colorlinks=true,linkcolor=blue,citecolor=blue,urlcolor=blue]{hyperref}
\usepackage{url}

\makeatletter
\def\@email#1#2{%
 \endgroup
 \patchcmd{\titleblock@produce}
  {\frontmatter@RRAPformat}
  {\frontmatter@RRAPformat{\produce@RRAP{*#1\href{mailto:#2}{#2}}}\frontmatter@RRAPformat}
  {}{}
}%
\makeatother

\begin{document}
\title{On Boltzmann Averaging in {\em Ab Initio} Thermodynamics}

\author{Hendrik H. Heenen}
\affiliation{Fritz-Haber-Institut der Max-Planck-Gesellschaft, Faradayweg 4-6, D-14195 Berlin, Germany}
\email{heenen@fhi.mpg.de}

\author{Karsten Reuter}
\affiliation{Fritz-Haber-Institut der Max-Planck-Gesellschaft, Faradayweg 4-6, D-14195 Berlin, Germany}
\email{reuter@fhi.mpg.de}

\begin{abstract}
{\em Ab initio} thermodynamics is a widespread, computationally efficient approach to predict the stable configuration of a surface in contact with a surrounding (gas or liquid) environment. In a prevalent realization of this approach, this stable configuration is simply equated with the structure in a considered candidate pool that exhibits the lowest surface free energy. Here we discuss the possibility to consider the thermal accessibility of competing, higher-energy configurations through Boltzmann averaging when the extended surface configurations and their energetics are computed within periodic boundary condition supercells. We show analytically that fully converged averages can be obtained with a candidate pool derived from exhaustive sampling in a surface unit-cell exceeding the system's correlation length. In contrast, averaging over a small pool of {\em ad hoc} assembled structures is generally ill-defined. Enumerations of a lattice-gas Hamiltonian model for on-surface oxygen adsorption at Pd(100) are employed to illustrate these considerations in a practical context.
\end{abstract}
\maketitle

\section{Introduction}

In {\em ab initio} thermodynamics we generally consider solid surface systems in equilibrium with an underlying solid bulk and a surrounding gas or liquid phase \cite{reuter2001,rogal2007,reuter2016,ringe2022,lee2024}. Within a grand-canonical ansatz, surface free energies of possible surface configurations are then computed as a function of the chemical potentials of the species in the environmental phase. For each set of values of the chemical potentials, this typically yields one configuration exhibiting the lowest surface free energy. In practical applications, this configuration is hitherto identified as the thermodynamically most stable one that the surface exhibits under the corresponding gas-phase conditions. The latter conclusion is, of course, strictly only correct in the limit of $T \rightarrow 0$\,K or if alternative surface configurations are thermally not accessible. At real surfaces, it is instead precisely this availability of numerous thermally accessible alternative configurations that stands behind crucial physics like surface inhomogeneity, defects and disorder at finite temperatures (even if the surface is fully equilibrated as assumed in {\em ab initio} thermodynamics). 

A prevailing realization of {\em ab initio} thermodynamics that we take as a starting point for our discussion often considers only a small (hand-selected) pool of surface configurations \cite{lee2024}. In this case, a sufficiently large surface free energy gap between the lowest and second lowest computed configuration is then qualitatively equated with "thermally not accessible", justifying the focus on the lowest-energy configuration. However, the situation is necessarily already fundamentally different at or close to chemical potential conditions where the energetic ordering of the lowest-energy configurations changes, i.e. at or close to a predicted phase boundary in the surface phase diagram \cite{rogal2007,reuter2016,reuter2003}. Within the $T \rightarrow 0$\,K approximation, this phase transition would be abrupt. Not least the implied non-physical discrete change of the surface configuration and its observables dictates to then go beyond the $T \rightarrow 0$ approximation and properly consider the thermal accessibility of competing surface configurations. In fact, this holds also more generally for cases of near energetic degeneracy of multiple alternative surface configurations, as will for instance likely arise when reduced computational costs e.g. through machine-learning surrogate models for the first-principles calculations \cite{bruix2019, bunting2023reactivity, du2023machine} lead to dramatically increased sizes of the considered pool of surface configurations.

In the following we discuss how and under which practical limitations such an account of thermal accessibility can be achieved through Boltzmann averaging. Through analytical derivations and numerical results for a lattice-gas Hamiltonian model for O adsorption at Pd(100) we demonstrate that Boltzmann averaging in {\em ab initio} thermodynamics with {\em ad hoc} created small candidate pools is generally ill-defined. Even if the candidate configurations instead result from a sufficiently systematic sampling, a major practical limitation derives from the size of the employed supercell. The surface area of this supercell must exceed the system's correlation length for the Boltzmann averages to be converged and meaningful.

\section{Formal Boltzmann averaging}

For the grand-canonical situation in {\em ab initio} thermodynamics, the prerequisite for Boltzmann averaging is the grand-canonical Boltzmann distribution which gives the probability with which the system will be in the various surface configurations at a given temperature $T$. With straightforward generalizations, we focus in the following on the case of a solid surface system formed of one species b in equilibrium with a surrounding gas phase formed of one species g. The two reservoirs are correspondingly the underlying solid bulk represented by a chemical potential $\mu_{\rm b}$, and the gas phase represented by a chemical potential $\mu_{\rm g}$. The surface system with a total surface area $A$ has microstates (aka surface configurations) $i$ with Gibbs free energies $G_i$ and total species numbers $N_{{\rm g},i}, N_{{\rm b},i}$. As derived for completeness in the appendix, the well known grand-canonical Boltzmann distribution for the probability $p_i$ of the system to be in microstate $i$ is then
\begin{equation}
\label{eq:Boltzdistr}
p_i \;=\; {\rm exp} \left( \frac{- G_i \,+\, \mu_{\rm g} N_{{\rm g},i} \,+\, \mu_{\rm b} N_{{\rm b},i}}{k_{\rm B}T} \right) / \mathcal{Z} \quad ,
\end{equation}
with the grand-canonical partition function
\begin{equation}
\label{eq:partitionfn}
\mathcal{Z} \;=\; \sum_j {\rm exp} \left( \frac{- G_j \,+\, \mu_{\rm g} N_{{\rm g},j} \,+\, \mu_{\rm b} N_{{\rm b},j}}{k_{\rm B}T} \right)
\end{equation}
and the Boltzmann constant $k_{\rm B}$.

With the common definition of the intensive, aka area-normalized, surface free energy
\begin{equation}
\gamma_i(\mu_{\rm g},\mu_{\rm b}) \;=\; \left( G_i \,-\, \mu_{\rm g} N_{{\rm g},i} \,-\, \mu_{\rm b} N_{{\rm b},i} \right) / A
\end{equation}
in {\em ab initio} thermodynamics \cite{rogal2007,reuter2016}, this can be rewritten as
\begin{equation}
\label{eq:gcbd}
p_i \;=\; {\rm exp} \left( \frac{- \gamma_i(\mu_{\rm g},\mu_{\rm b})A}{k_{\rm B}T} \right) / \mathcal{Z} \quad ,
\end{equation}
with
\begin{equation}
\label{eq:gcpf}
\mathcal{Z} \;=\; \sum_j {\rm exp} \left( \frac{- \gamma_j(\mu_{\rm g},\mu_{\rm b})A}{k_{\rm B}T} \right)
\quad .
\end{equation}
With this distribution, thermal averages at temperature $T$ of observables $O$ are then obtained as 
\begin{equation}
\label{eq:observ}
\langle O \rangle \;=\; \sum_i p_i O_i \quad .
\end{equation}

Formally, an extension beyond the prevalent $T \rightarrow 0$\,K approximation in {\em ab initio} thermodynamics is thus straightforward. As long as all microstates $i$ of the system, i.e. all possible surface configurations, are known, a consideration of thermal accessibility through Boltzmann averaging can be achieved through eq.~(\ref{eq:observ}). Note that this statement reflects the inherent workflow in the above described practical realization of {\em ab initio} thermodynamics, where first possible surface configurations $i$ with different stoichiometries are identified (either by domain knowledge, enumeration or some form of uniform sampling), and their relative energetic ordering and concomitant relevance is only subsequently evaluated within the grand-canonical scheme. This is thus fundamentally different to grand-canonical Boltzmann sampling schemes, where the microstates $i$ of the system are directly sampled according to a Boltzmann distribution in the first place. In this case, the Boltzmann averaging in the then obtained ensemble of $M$ microstates $\{i'\}$ is merely 
\begin{equation}
\label{eq:gcsamp}
\langle O \rangle \;=\; \frac{1}{M} \sum_{i' = 0}^M O_{i'} \quad .
\end{equation}

\section{Boltzmann averaging in supercells}

In the first-principles calculations underlying {\em ab initio} thermodynamics, extended surfaces are predominantly described within periodic boundary condition supercells, i.e.\ with surface slabs of finite thickness and a surface unit-cell area $A_{\rm surf}$. While derived generically, the above grand-canonical Boltzmann distribution still holds for such a supercell setup as long as $A_{\rm surf}$ exceeds the correlation length of the surface configurations of the system under study. In this case, the probability for the surface to be in any microstate $i$ within the area $A_{\rm surf}$ is independent of the probability with which the system is in other microstates $j$ anywhere else on the surface. Correspondingly, the Boltzmann average of any intensive observable is then also independent of the $A_{\rm surf}$ chosen in the practical calculations (or additive in case of extensive observables).

The latter statement is readily proven by splitting the system into two subsystems A and B with subareas $A_{\rm surf,A}$ and $A_{\rm surf,B}$, respectively $(A_{\rm surf,A} + A_{\rm surf,B} = A_{\rm surf})$. The splitting is thereby completely arbitrary, as long as the two subareas are each still larger than the correlation length. In this case, the ensuing independence of the two subsystems renders their grand-canonical partition functions multiplicative
\begin{equation}
\label{eq:Z_AB}
\mathcal{Z} \;=\; \mathcal{Z}_{\rm A}\mathcal{Z}_{\rm B} \quad .
\end{equation}
Using eq.~(\ref{eq:gcbd}), it is then straightforward to prove that the probability to find microstate $i$ (with $\gamma_i$) simultaneously in both subsystems is the same as the probability to find it in the full system
\begin{equation}
\label{eq:pi_AB}
p_{i({\rm A})} p_{i({\rm B})} \;=\; p_i \quad .
\end{equation}

Here it is important to note that this probability is, of course, smaller than the probability to find the microstate in any one of the smaller subsystems, i.e. generally $p_i < p_{i({\rm A})}$ and $p_i < p_{i({\rm B})}$. The likelihood to find a higher energy configuration extending over a larger and larger domain on the surface obviously does become smaller. However, it is equally crucial to realize that the number of microstates in the full system is larger than in each of the two subsystems. Most accessibly, this can e.g. be seen when considering a model surface system with only two surface configurations 0 and 1 with surface free energies $\gamma_0$ and $\gamma_1$, respectively. Splitting the surface area into two halves, the sum over microstates thus involves only these two microstates 0 and 1. However, beyond the correlation length this sum in the full system involves four states, i.e. beyond the states 00 and 11 (where the total surface is homogeneously covered with either configuration) also the states 01 and 10 (where half of the surface is in either configuration).

Generally, we thus have
\begin{equation}
\label{eq:sum_states}
\sum_i \;=\; \sum_{i({\rm A})} \sum_{i({\rm B})} \quad .
\end{equation}
Since beyond the correlation length also any extensive observable is additive, i.e. $O_{i} = O_{i({\rm A})} + O_{i({\rm B})}$, it is again straightforward to prove that
\begin{eqnarray}
\label{eq:av_add}
\langle O_i \rangle &=& \sum_i p_i O_i \;=\; \sum_{i({\rm A})} \sum_{i({\rm B})} p_{i({\rm A})} p_{i({\rm B})} (O_{i({\rm A})} + O_{i({\rm B})}) \\ \nonumber
&=& \sum_{i({\rm A})} p_{i({\rm A})} O_{i({\rm A})} \;+\; \sum_{i({\rm B})} p_{i({\rm B})} O_{i({\rm B})} \\ \nonumber
&=& \langle O_{i({\rm A})} \rangle \;+\; \langle O_{i({\rm B})} \rangle \quad ,
\end{eqnarray}
where we have used the normalization of the probabilities within each subsystem, i.e. $\sum_{i({\rm A})} p_{i({\rm A})} = 1$ and $\sum_{i({\rm B})} p_{i({\rm B})} = 1$. This thus shows analytically that the Boltzmann average of any extensive observable is additive as long as the surface unit-cell employed in the supercell simulations exceeds the correlation length on the surface. Analogously, one can prove that for any intensive observable, the Boltzmann average is then independent of the chosen surface unit-cell size. To this end, we note in passing that the system's correlation length can trivially be exceeded if the actual difference in surface configurations is restricted to a finite sub-part on the extended surface like prominently in the case of a small cluster on a static support \cite{kumari2023nature, zhang2020ensembles}. As long as the surface unit-cell is chosen large enough to decouple the finite part (cluster) from its periodic images, the correlation length is exceeded and Boltzmann averaging can straightforwardly be performed. Analytically, this already results because any additional contributions to the $G_i$ from a varying surface area beyond the finite part are equal in all configurations and thus cancel out in eqs.~(\ref{eq:Boltzdistr}) and (\ref{eq:partitionfn}).

\section{Corollaries for practical calculations}

From the analytical considerations up to this point, the straightforward best practice to consider thermal accessibility in {\em ab initio} thermodynamics through Boltzmann averaging is thus to use a supercell with $A_{\rm surf}$ exceeding the system's correlation length and to achieve an exhaustive consideration of the system's surface configurations within this cell. Such an exhaustive consideration could conceptually e.g.\ be achieved by systematic enumeration or a sufficiently excessive uniform sampling. At an in general {\em a priori} unknown correlation length, this would furthermore also involve systematic convergence tests for $\langle O \rangle$ with increasing cell size.

In real applications, this conceptual best practice is unlikely tractable though. Probably, already an extensive, but unlikely fully exhaustive sampling within one large supercell may be at the edge of what is computationally doable even when employing efficient machine-learning surrogate models for the first-principles calculations. Beyond sampling deficiencies, it is then primarily potential finite-size effects due to an insufficiently large $A_{\rm surf}$ that will affect the computed thermal averages. To this end, we note that one way to address sampling deficiencies could be to switch to grand-canonical Boltzmann sampling \cite{karlberg2007cyclic, zhou2022ab, zhou2019determining}, i.e. to adapt the workflow as already described above. Instead of first uniformly sampling the system's microstates and then Boltzmann averaging their thermal contributions to observables as in eq.~(\ref{eq:observ}), molecular dynamics and/or Monte Carlo based Boltzmann sampling would directly reflect the thermal probability of the different microstates and reduce the observable averaging to the simple eq.~(\ref{eq:gcsamp}).

While such Boltzmann sampling may prove to be numerically more efficient than uniform sampling, it would still suffer from finite-size effects if the chosen $A_{\rm surf}$ is too small. The largest contributor to such finite-size effects are surface configurations with structures that do not commensurably fit into the employed simulation cell. They will then either be completely absent in the sampled candidate pool or only be present in strained versions or with certain defects like domain boundaries. At tractable surface unit-cells the latter strains or defects likely impose considerable energy penalties, which then significantly affect the probability computed for such configurations and their corresponding contribution to the thermal averages (both in uniform or Boltzmann sampling approaches). 

In principle, one could consider mitigating these limitations in the uniform sampling or enumeration approach by extension to multiple surface unit-cells of different shapes and sizes, and then considering all correspondingly obtained microstates and their surface free energies in the Boltzmann average at one chosen $A$ in eq.~(\ref{eq:gcbd}), see below. At an extreme, this could even extend to explicitly computing certain known ordered surface structures in their corresponding commensurate cells and including them as well. The minimum prerequisite here would be to use fully consistent computational settings, and in particular compatible k-meshes, for all employed supercells as already imperative for traditional small candidate-pool {\em ab initio} thermodynamics. However, such a combination of microstates from different simulation cells runs the danger of double counting surface configurations and is prone to generating an unbalanced pool with different structures affected differently by varying degrees of strain. In our view, the better controllable alternative is therefore to run the sampling in one cell with as large an $A_{\rm surf}$ as possible, perform the averaging over the resulting ensemble of surface configurations and find ways to at least qualitatively assess potential (hopefully residual) finite-size effects.

From this perspective, it is finally worthwhile to comment on the possibility of Boltzmann averaging in traditional small-pool {\em ab initio} thermodynamics. Here, the considered candidate structures are typically fully ordered and computed in different surface unit-cells that each perfectly match the structure's periodicity. While this eliminates strain effects, this approach can be seen as a particularly poor uniform sampling in a gigantic surface unit-cell that commensurably encompasses all of the different surface unit-cells of the considered candidate structures. As this small candidate pool obviously lacks the most likely low-energy (and thus thermally accessible) surface modifications in form of (point) defects and/or phase coexistence, Boltzmann averaging makes in principle no sense. Due to the gross sampling deficiencies, any attempted Boltzmann averaging will also explicitly depend on the $A$ chosen in eq.~(\ref{eq:gcbd}) and is as such ill-defined. In fact, the latter holds equally for the idea to combine microstates enumerated or sampled in different surface unit-cells as discussed above, even if the considered candidate pools are much larger. 

If a Boltzmann-averaging in small-pool {\em ab initio} thermodynamics is at all attempted, the $A$ in eqs.~(\ref{eq:gcbd}-\ref{eq:gcpf}) with which all normalized surface free energies are multiplied should probably be chosen as small as possible. Natural choices would for instance be the area of a $(1 \times 1)$ surface unit-cell of the truncated bulk crystal or one could rewrite the $\gamma_i$ into formation energies per one surface species. In fact, such choices have been made recurrently in the literature, but, unfortunately, without any critical discussion on the why and on how this affects the obtained results. To state this therefore quite clearly: even for a minimal $A$ the derived Boltzmann average will, of course, still depend on this area choice (and suffer from the unsystematic candidate pool). As such, any obtained results are arbitrary. The only thing achieved with a minimal $A$ is that the grand-canonical Boltzmann distribution implicitly considers at least mixed configurations that under the choice of a larger $A$ would have to be explicitly added to the candidate pool. In other words, if one chooses a $(1 \times 1)$ surface unit-cell as area, the Boltzmann distribution implicitly considers all inhomogeneous surface configurations that can combinatorially be formed out of $(1 \times 1)$ domains of the structures in the candidate pool. It does so under the complete neglect of any costs for the domain boundaries in them though. 

So, while maximizing the candidate pool that is implicitly accounted for in the Boltzmann distribution, this created pool is likely a rough approximation. Of course, it isn't if the boundary costs are indeed negligible. In this case, the minimal area chosen also exceeds the correlation length of the system and the small-pool approach can be most efficient in achieving what a laborious sampling or enumeration would do in larger cells. A classic example for this is the adsorption of a gas-phase species at a defined site of a crystalline surface with negligible lateral interactions between the adsorbed species. In this case, a pool of two candidate structures in form of the clean surface with $G_{\rm clean}$ and a surface with a $(1 \times 1)$-coverage of the adsorbate species with $G_{\rm (1 \times 1)}$ is fully sufficient. Choosing $A$ to be the area of the $(1 \times 1)$ surface unit-cell and evaluating eqs.~(\ref{eq:gcbd}-\ref{eq:observ}), leads to a Boltzmann averaged coverage $\langle \theta \rangle$ as a function of the gas-phase chemical potential $\mu_{\rm g}$ that correctly produces the expected Langmuir adsorption isotherm \cite{masel_1996}
\begin{equation}
\langle \theta \rangle \;=\; \frac{1}{1 \;+\; {\rm exp} \left( \frac{\Delta G_{\rm ad} - \mu_{\rm g}}{k_{\rm B}T} \right)} \quad ,
\end{equation}
where $\Delta G_{\rm ad} = (G_{\rm (1 \times 1)} - G_{\rm clean})$ is the adsorption free energy difference.

\section{Numerical results: O at Pd(100)}

\begin{figure*}[ht!]
\centering\includegraphics[width=0.9\textwidth]{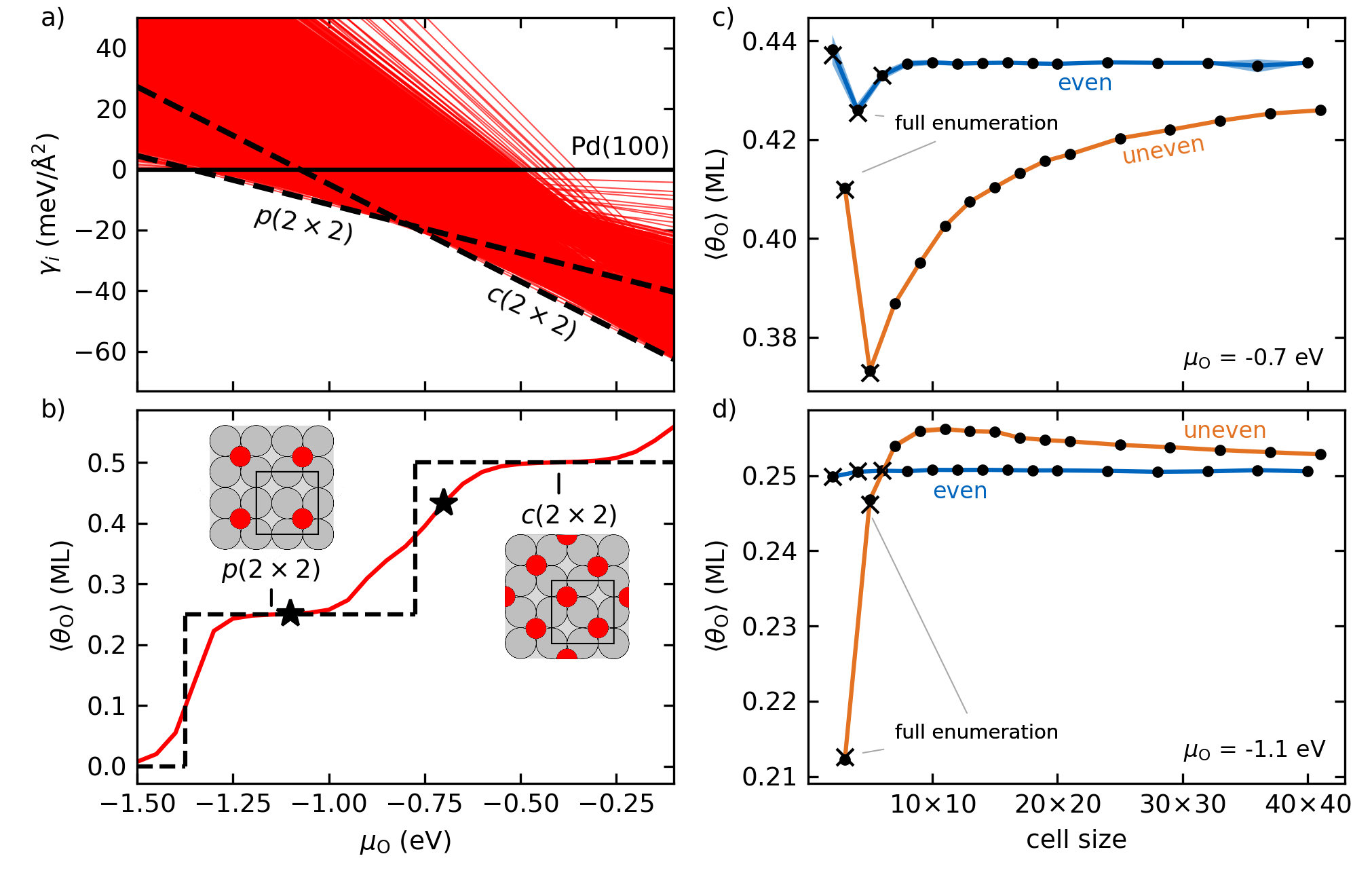}
\caption{Lattice-gas Hamiltonian described low-coverage adsorption regime of on-surface O at Pd(100). a) Surface free energies $\gamma_i$ as a function of the O chemical potential $\mu_{\rm O}$ of the two ordered structures, $p(2 \times 2)$-O at 0.25\,ML and $c(2 \times 2)$-O at 0.5\,ML (dashed thick black lines), as well as of all configurations obtained from full enumeration in a $(6 \times 6)$ surface unit-cell (solid thin red lines). b) Corresponding coverage curve $\langle \theta_{\rm O} \rangle$ as obtained within the $T \rightarrow 0$\,K approximation (dashed black line) and from Boltzmann averaging at 600\,K (solid red line). The insets show top views of the two ordered structures (O small red spheres, Pd large gray spheres). c) and d) Convergence behavior of $\langle \theta_{\rm O} \rangle$ with surface unit-cell size employed in the sampling for two specific values $\mu_{\rm O}$ (marked with $\star$ in b)). $\times$ data points in the smaller unit-cells are obtained by full enumeration, $\circ$ data points by Boltzmann sampling, see text.}
\label{fig1}
\end{figure*}

We finally illustrate the above considerations for a simple, but representative system. Specifically, this concerns the on-surface adsorption of O at a Pd(100) surface, as described through a first-principles lattice-gas Hamiltonian (LGH). The LGH includes pair, trio and quattro lateral interactions derived by Zhang {\em et al.} \cite{zhang_2007}, and faithfully reproduces the low-coverage adsorption regime as characterized by two experimentally reported ordered phases, a $p(2 \times 2)$-O structure at 0.25\,monolayer (ML) and a $c(2 \times 2)$-O structure at 0.5 ML, both with O adsorbed in the on-surface fourfold hollow sites. With details on the numerical simulations provided in the appendix, Fig.~\ref{fig1}a shows the LGH-computed surface free energies of these two ordered structures as a function of the oxygen chemical potential $\mu_{\rm O}$. Due to the increasing O-content and thus steeper slopes of the corresponding $\gamma_i$ lines with increasing $\mu_{\rm O}$, first the $p(2 \times 2)$-O becomes most stable above $\mu_{\rm O} = - 1.38$\,eV and then the $c(2 \times 2)$-O at $\mu_{\rm O} = -0.78$\,eV. Within the $T \rightarrow 0$\,K approximation, the derived coverage curve shown in Fig.~\ref{fig1}b correspondingly has a step-function shape with two plateaus at 0.25\,ML and 0.5\,ML. 

Also included in the figure is the average coverage $\langle \theta_{\rm O} \rangle$ at 600\,K as obtained through eq.~(\ref{eq:observ}) and considering all surface configurations $i$ enumerated in a $(6 \times 6)$ surface unit-cell. As expected, the Boltzmann averaging eliminates the unphysical step-changes in the coverage curve and replaces them with smooth gradual transitions. The right panels in Fig.~\ref{fig1}c and d finally show the convergence behavior of $\langle \theta_{\rm O} \rangle$ with surface unit-cell size at two selected chemical potentials, one within the plateau of the $p(2 \times 2)$ and one within the smoothened transition from the $p(2 \times 2)$ to the $c(2 \times 2)$ phase. In the smaller surface unit-cells we fully enumerate all possible microstates aka adsorption configurations and then consider them in the Boltzmann averaging according to eq.~(\ref{eq:observ}). Even though this number of microstates rises steeply, amounting to 68,719,476,736 states in the $(6 \times 6)$ cell, the computed average coverage converges slowly and exhibits a zig-zag behavior between even and odd sized cells. The latter derives from the inability to accommodate the ordered $p(2 \times 2)$ (or $c(2 \times 2)$) structures in odd-sized cells, highlighting the finite-size effects described in general terms above. As full enumeration eventually becomes cumbersome (intractable), we switch to Boltzmann sampling via grand-canonical Monte Carlo (GCMC) simulations for the larger surface unit-cells. The identical results obtained for the $(2 \times 2)$ to $(6 \times 6)$ cells by full enumeration and Boltzmann averaging via eq.~(\ref{eq:observ}) on the one hand and GCMC sampling and averaging via eq.~(\ref{eq:gcsamp}) on the other demonstrate numerically the expected equivalence of these two approaches. The results for the larger cells in Fig.~\ref{fig1}c and d then confirm that the Boltzmann averages can indeed be converged. However, even for this simple system, quite large surface unit-cells are required to reach and exceed the concomitant correlation length.

\section{Conclusions}

The present analysis shows that Boltzmann averaging as a means to overcome the prevalent $T \rightarrow 0$\,K approximation in {\em ab initio} thermodynamics is viable in principle. In practice, it cannot reliably be performed on an {\em ad hoc} assembled (small) pool of (ordered) surface configurations though, but requires systematic sampling approaches. Uniform sampling and a subsequent Boltzmann averaging thereby maintains the classic workflow of {\em ab initio} thermodynamics, whereas direct Boltzmann sampling offers an equivalent ansatz that is potentially numerically more efficient. More problematic is arguably to achieve convergence with respect to the surface unit-cell employed in the sampling. We generally expect finite-size effects even in the larger cell sizes tractable with efficient machine-learning interatomic potentials as surrogate for the first-principles calculations. Obtaining reliable estimates for these finite-size effects and finding means to correct for them is in our view the crucial next methodological frontier if Boltzmann averaging is indeed to be pursued in the {\em ab initio} thermodynamics context to obtain quantitative averages for observables at finite temperatures. If this can be achieved, it will open the door to systematically account for surface inhomogeneity and disorder within this widespread approach.

\begin{acknowledgments}
We acknowledge support from the German Research Foundation for funding through the DFG Cluster of Excellence e-conversion EXC 2089/1 and the Max Planck Computing and Data Facility (MPCDF) for providing computing time.
\end{acknowledgments}

\section*{Author Declarations}

\subsection*{Conflict of Interest}

The authors have no conflicts to disclose.

\subsection*{Author Contributions}

K.R.\ conceptualized and prepared the original draft. H.H.H.\ developed methodology and software for the numerical results, and edited the draft.

\section*{Data Availability}

The code and data for the numerical results is available in the GitHub repository at \url{https://github.com/hheenen/LGH_icet_PdOads} and will be made public after publication of the article.

\appendix
\section{Derivation of the grand-canonical Boltzmann distribution}

For completeness we rederive the known grand-canonical Boltzmann distribution for the surface system considered in the main text. The probability for each microstate in the grand-canonical setup is $p_i$. The Gibbs-Shannon entropy correspondingly reads
\begin{equation}
\label{eq:shannon}
S \;=\; - k_{\rm B} \sum_i \left( p_i {\rm ln}(p_i) \right) \quad .
\end{equation}

As constraints we have the normalization of the probabilities
\begin{equation}
\label{eq:norm}
\sum_i p_i \;=\; 1 \quad ,
\end{equation}
as well as the average Gibbs free energy
\begin{equation}
\label{eq:aveng}
\langle G \rangle \;=\; \sum_i p_i \, G_i \quad 
\end{equation}
and the average species numbers
\begin{equation}
\langle N_{\rm g} \rangle \;=\; \sum_i p_i \, N_{{\rm g},i} \quad  
\end{equation}
and
\begin{equation}
\label{eq:avpart}
\langle N_{\rm b} \rangle \;=\; \sum_i p_i \, N_{{\rm b},i} \quad .
\end{equation}

Entering the enforced constraints via Langrange multipliers $\alpha$, $\beta$, $\gamma$, $\delta$, the entropy reads
\begin{eqnarray}
\label{eq:entropy}
S' &=& - k_{\rm B} \sum_i \left( p_i {\rm ln}(p_i) \right) \;+\; \alpha \left(\sum_i p_i - 1\right) \;+ \\ \nonumber
   & &  \beta \left(\sum_i p_i \, G_i  - \langle G \rangle\right) \;+\; \gamma \left(\sum_i p_i \, N_{{\rm g},i} - \langle N_{\rm g} \rangle\right) \;+ \\ \nonumber
	 & &  \delta \left(\sum_i p_i \, N_{{\rm b},i} - \langle N_{\rm b} \rangle\right) \quad .
\end{eqnarray}
The most probable distribution maximizes this entropy at chemical equilibrium and we correspondingly set the derivative of eq.~(\ref{eq:entropy}) with respect to $p_i$ to zero:
\begin{equation}
\partial S'/\partial p_i \;=\; - k_{\rm B} \left( 1 + {\rm ln}(p_i) \right) \;+\; \alpha \;+\; \beta G_i \;+\; \gamma N_{{\rm g},i} \;+\;
\delta N_{{\rm b},i} \;=\; 0 \\ \quad .
\end{equation}
Solving for $p_i$ then leads to
\begin{equation}
p_i \;=\; {\rm exp} \left( \frac{-\alpha \,-\, \beta G_i \,-\, \gamma N_{{\rm g},i} \,-\, \delta N_{{\rm b},i}}{k_{\rm B}} \right) \quad .
\end{equation}
Entering this into the normalization constraint, eq.~(\ref{eq:norm}), allows to identify
\begin{equation}
{\rm exp} \left( \frac{\alpha}{k_{\rm B}} \right) \;=\; \sum_i {\rm exp} \left( \frac{-\beta G_i \,-\, \gamma N_{{\rm g},i} \,-\, \delta N_{{\rm b},i}}{k_{\rm B}} \right) \quad ,
\end{equation}
so that we arrive at
\begin{eqnarray}
p_i &=& \frac{ {\rm exp} \left( \frac{-\beta G_i \,-\, \gamma N_{{\rm g},i} \,-\, \delta N_{{\rm b},i}}{k_{\rm B}} \right)}{\sum_j {\rm exp} \left( \frac{-\beta G_j \,-\, \gamma N_{{\rm g},j} \,-\, \delta N_{{\rm b},j}}{k_{\rm B}} \right)} \\ \nonumber
    &=& {\rm exp} \left( \frac{-\beta G_i \,-\, \gamma N_{{\rm g},i} \,-\, \delta N_{{\rm b},i}}{k_{\rm B}} \right) / \mathcal{Z} \quad .
\end{eqnarray}
where we have introduced the grand-canonical partition function
\begin{equation}
\mathcal{Z} \;=\; \sum_j {\rm exp} \left( \frac{-\beta G_j \,-\, \gamma N_{{\rm g},j} \,-\, \delta N_{{\rm b},j}}{k_{\rm B}} \right)
\end{equation}
in the second equation.

We now enter this into the Gibbs-Shannon entropy definition, eq.~(\ref{eq:shannon}), and exploit the constraining eqs.~({\ref{eq:aveng} - \ref{eq:avpart}),
\begin{eqnarray}
S &=& - k_{\rm B} \sum_i p_i {\rm ln}(p_i) \\ \nonumber
  &=& - k_{\rm B} \sum_i p_i {\rm ln} \left(  \frac{ {\rm exp} \left( \frac{-\beta G_i \,-\, \gamma N_{{\rm g},i} \,-\, \delta N_{{\rm b},i}}{k_{\rm B}} \right)}{\mathcal Z} \right) \\ \nonumber
  &=& - k_{\rm B} \sum_i p_i \left( \frac{-\beta G_i \,-\, \gamma N_{{\rm g},i} \,-\, \delta N_{{\rm b},i}}{k_{\rm B}} \;-\; {\rm ln}\mathcal{Z} \right) \\ \nonumber
	&=& \beta \sum_i p_i G_i \;+\; \gamma \sum_i p_i N_{{\rm g},i} \;+\; \delta \sum_i p_i N_{{\rm b},i} \;+\; k_{\rm B} {\rm ln}\mathcal{Z} \\ \nonumber
	&=& \beta \langle G \rangle \;+\; \gamma \langle N_{\rm g} \rangle \;+\; \delta \langle N_{\rm b} \rangle \;+\; k_{\rm B} {\rm ln}\mathcal{Z} \quad .
	\end{eqnarray}
Using the thermodynamic relations for the partial derivatives of the entropy
\begin{eqnarray}
\left(\frac{\partial S}{\partial \langle G \rangle}\right)_{V,N_{\rm g},N_{\rm b}} &=& \frac{1}{T} \\ \nonumber
\left(\frac{\partial S}{\partial \langle N_{\rm g} \rangle}\right)_{V,E,N_{\rm b}} &=& -\frac{\mu_{\rm g}}{T} \\ \nonumber
\left(\frac{\partial S}{\partial \langle N_{\rm b} \rangle}\right)_{V,E,N_{\rm g}} &=& -\frac{\mu_{\rm b}}{T} \quad ,
\end{eqnarray}
thus allows to identify the Lagrange multipliers as
\begin{eqnarray}
\beta &=& \frac{1}{T} \\ \nonumber
\gamma &=& -\frac{\mu_{\rm g}}{T} \\ \nonumber
\delta &=& -\frac{\mu_{\rm b}}{T} \quad .
\end{eqnarray}
Correspondingly, we arrive at the final form of the grand-canonical Boltzmann distribution for the probabilities of the system to be in microstate $i$
\begin{equation}
p_i \;=\; {\rm exp} \left( \frac{- G_i \,+\, \mu_{\rm g} N_{{\rm g},i} \,+\, \mu_{\rm b} N_{{\rm b},i}}{k_{\rm B}T} \right) / \mathcal{Z} \quad ,
\end{equation}
with the grand-canonical partition function
\begin{equation}
\mathcal{Z} \;=\; \sum_j {\rm exp} \left( \frac{- G_j \,+\, \mu_{\rm g} N_{{\rm g},j} \,+\, \mu_{\rm b} N_{{\rm b},j}}{k_{\rm B}T} \right)
\quad .
\end{equation}

\section{Details of numerical simulations}

The LGH model by Zhang \textit{et al.} \cite{zhang_2007} provides O adsorption energies referenced against the clean Pd(100) surface and an isolated oxygen molecule at 0\,K and neglecting zero-point energy. The latter is therefore also the zero reference for the $\mu_{\rm O}$ scale employed in Fig.~\ref{fig1}. For reasons of numerical efficiency, we refitted the LGH via a cluster expansion formalism with an equivalent interaction cut-off using the icet code \cite{aangqvist2019icet}. This refitting process involved 27 original data points from ref.~\cite{zhang_2007} and 150 additional data points of randomly generated $\theta_{\rm O} \leq 0.5$ configurations in $(2 \times 2)$ to $(4 \times 4)$ supercells evaluated by our own reimplementation of the original LGH. We added the latter data points to ensure improved accuracy in the low-coverage regime which was otherwise enforced in the original LGH by a fixed on-site energy. The refitted model accurately reproduces the original data with a cross-validation score of 8\,meV and a root-mean-square error of 65\,meV. The GCMC simulations employed 2000 MC passes per site for equilibration and 10000 MC passes per site for sampling, at which point we found the coverages fully converged to within 0.01\,ML.

\bibliography{JCP_Heenen.bib}

\begin{thebibliography}{17}%
\makeatletter
\providecommand \@ifxundefined [1]{%
 \@ifx{#1\undefined}
}%
\providecommand \@ifnum [1]{%
 \ifnum #1\expandafter \@firstoftwo
 \else \expandafter \@secondoftwo
 \fi
}%
\providecommand \@ifx [1]{%
 \ifx #1\expandafter \@firstoftwo
 \else \expandafter \@secondoftwo
 \fi
}%
\providecommand \natexlab [1]{#1}%
\providecommand \enquote  [1]{``#1''}%
\providecommand \bibnamefont  [1]{#1}%
\providecommand \bibfnamefont [1]{#1}%
\providecommand \citenamefont [1]{#1}%
\providecommand \href@noop [0]{\@secondoftwo}%
\providecommand \href [0]{\begingroup \@sanitize@url \@href}%
\providecommand \@href[1]{\@@startlink{#1}\@@href}%
\providecommand \@@href[1]{\endgroup#1\@@endlink}%
\providecommand \@sanitize@url [0]{\catcode `\\12\catcode `\$12\catcode
  `\&12\catcode `\#12\catcode `\^12\catcode `\_12\catcode `\%12\relax}%
\providecommand \@@startlink[1]{}%
\providecommand \@@endlink[0]{}%
\providecommand \url  [0]{\begingroup\@sanitize@url \@url }%
\providecommand \@url [1]{\endgroup\@href {#1}{\urlprefix }}%
\providecommand \urlprefix  [0]{URL }%
\providecommand \Eprint [0]{\href }%
\providecommand \doibase [0]{https://doi.org/}%
\providecommand \selectlanguage [0]{\@gobble}%
\providecommand \bibinfo  [0]{\@secondoftwo}%
\providecommand \bibfield  [0]{\@secondoftwo}%
\providecommand \translation [1]{[#1]}%
\providecommand \BibitemOpen [0]{}%
\providecommand \bibitemStop [0]{}%
\providecommand \bibitemNoStop [0]{.\EOS\space}%
\providecommand \EOS [0]{\spacefactor3000\relax}%
\providecommand \BibitemShut  [1]{\csname bibitem#1\endcsname}%
\let\auto@bib@innerbib\@empty
\bibitem [{\citenamefont {Reuter}\ and\ \citenamefont
  {Scheffler}(2001)}]{reuter2001}%
  \BibitemOpen
  \bibfield  {author} {\bibinfo {author} {\bibfnamefont {K.}~\bibnamefont
  {Reuter}}\ and\ \bibinfo {author} {\bibfnamefont {M.}~\bibnamefont
  {Scheffler}},\ }\bibfield  {title} {\enquote {\bibinfo {title} {Composition,
  structure, and stability of ruo$_2$(110) as a function of oxygen pressure},}\
  }\href@noop {} {\bibfield  {journal} {\bibinfo  {journal} {Phys. Rev. B}\
  }\textbf {\bibinfo {volume} {65}},\ \bibinfo {pages} {035406} (\bibinfo
  {year} {2001})}\BibitemShut {NoStop}%
\bibitem [{\citenamefont {Rogal}\ and\ \citenamefont
  {Reuter}(2007)}]{rogal2007}%
  \BibitemOpen
  \bibfield  {author} {\bibinfo {author} {\bibfnamefont {J.}~\bibnamefont
  {Rogal}}\ and\ \bibinfo {author} {\bibfnamefont {K.}~\bibnamefont {Reuter}},\
  }\bibfield  {title} {\enquote {\bibinfo {title} {Ab initio atomistic
  thermodynamics for surfaces: A primer},}\ }\href@noop {} {\bibfield
  {journal} {\bibinfo  {journal} {Experiment, modeling and simulation of
  gas-surface interactions for reactive flows in hypersonic flights}\ }\textbf
  {\bibinfo {volume} {14}},\ \bibinfo {pages} {2--1} (\bibinfo {year}
  {2007})}\BibitemShut {NoStop}%
\bibitem [{\citenamefont {Reuter}(2016)}]{reuter2016}%
  \BibitemOpen
  \bibfield  {author} {\bibinfo {author} {\bibfnamefont {K.}~\bibnamefont
  {Reuter}},\ }\bibfield  {title} {\enquote {\bibinfo {title} {Ab initio
  thermodynamics and first-principles microkinetics for surface catalysis},}\
  }\href@noop {} {\bibfield  {journal} {\bibinfo  {journal} {Catal. Lett.}\
  }\textbf {\bibinfo {volume} {146}},\ \bibinfo {pages} {541--563} (\bibinfo
  {year} {2016})}\BibitemShut {NoStop}%
\bibitem [{\citenamefont {Ringe}\ \emph {et~al.}(2022)\citenamefont {Ringe},
  \citenamefont {H{\"o}rmann}, \citenamefont {Oberhofer},\ and\ \citenamefont
  {Reuter}}]{ringe2022}%
  \BibitemOpen
  \bibfield  {author} {\bibinfo {author} {\bibfnamefont {S.}~\bibnamefont
  {Ringe}}, \bibinfo {author} {\bibfnamefont {N.}~\bibnamefont {H{\"o}rmann}},
  \bibinfo {author} {\bibfnamefont {H.}~\bibnamefont {Oberhofer}},\ and\
  \bibinfo {author} {\bibfnamefont {K.}~\bibnamefont {Reuter}},\ }\bibfield
  {title} {\enquote {\bibinfo {title} {Implicit solvation methods for catalysis
  at electrified interfaces},}\ }\href@noop {} {\bibfield  {journal} {\bibinfo
  {journal} {Chem. Rev.}\ }\textbf {\bibinfo {volume} {122}},\ \bibinfo {pages}
  {10777} (\bibinfo {year} {2022})}\BibitemShut {NoStop}%
\bibitem [{\citenamefont {Lee}\ and\ \citenamefont {Soon}(2024)}]{lee2024}%
  \BibitemOpen
  \bibfield  {author} {\bibinfo {author} {\bibfnamefont {T.}~\bibnamefont
  {Lee}}\ and\ \bibinfo {author} {\bibfnamefont {A.}~\bibnamefont {Soon}},\
  }\bibfield  {title} {\enquote {\bibinfo {title} {The rise of ab initio
  surface thermodynamics},}\ }\href@noop {} {\bibfield  {journal} {\bibinfo
  {journal} {Nat. Catal.}\ }\textbf {\bibinfo {volume} {7}},\ \bibinfo {pages}
  {4--6} (\bibinfo {year} {2024})}\BibitemShut {NoStop}%
\bibitem [{\citenamefont {Reuter}\ and\ \citenamefont
  {Scheffler}(2003)}]{reuter2003}%
  \BibitemOpen
  \bibfield  {author} {\bibinfo {author} {\bibfnamefont {K.}~\bibnamefont
  {Reuter}}\ and\ \bibinfo {author} {\bibfnamefont {M.}~\bibnamefont
  {Scheffler}},\ }\bibfield  {title} {\enquote {\bibinfo {title}
  {First-principles atomistic thermodynamics for oxidation catalysis: Surface
  phase diagrams and catalytically interesting regions},}\ }\href@noop {}
  {\bibfield  {journal} {\bibinfo  {journal} {Phys. Rev. Lett.}\ }\textbf
  {\bibinfo {volume} {90}},\ \bibinfo {pages} {046103} (\bibinfo {year}
  {2003})}\BibitemShut {NoStop}%
\bibitem [{\citenamefont {Bruix}\ \emph {et~al.}(2019)\citenamefont {Bruix},
  \citenamefont {Margraf}, \citenamefont {Andersen},\ and\ \citenamefont
  {Reuter}}]{bruix2019}%
  \BibitemOpen
  \bibfield  {author} {\bibinfo {author} {\bibfnamefont {A.}~\bibnamefont
  {Bruix}}, \bibinfo {author} {\bibfnamefont {J.}~\bibnamefont {Margraf}},
  \bibinfo {author} {\bibfnamefont {M.}~\bibnamefont {Andersen}},\ and\
  \bibinfo {author} {\bibfnamefont {K.}~\bibnamefont {Reuter}},\ }\bibfield
  {title} {\enquote {\bibinfo {title} {First-principles based multiscale
  modeling of heterogeneous catalysis},}\ }\href@noop {} {\bibfield  {journal}
  {\bibinfo  {journal} {Nature Catal.}\ }\textbf {\bibinfo {volume} {2}},\
  \bibinfo {pages} {659} (\bibinfo {year} {2019})}\BibitemShut {NoStop}%
\bibitem [{\citenamefont {Bunting}\ \emph {et~al.}(2023)\citenamefont
  {Bunting}, \citenamefont {Wodaczek}, \citenamefont {Torabi},\ and\
  \citenamefont {Cheng}}]{bunting2023reactivity}%
  \BibitemOpen
  \bibfield  {author} {\bibinfo {author} {\bibfnamefont {R.~J.}\ \bibnamefont
  {Bunting}}, \bibinfo {author} {\bibfnamefont {F.}~\bibnamefont {Wodaczek}},
  \bibinfo {author} {\bibfnamefont {T.}~\bibnamefont {Torabi}},\ and\ \bibinfo
  {author} {\bibfnamefont {B.}~\bibnamefont {Cheng}},\ }\bibfield  {title}
  {\enquote {\bibinfo {title} {Reactivity of single-atom alloy nanoparticles:
  modeling the dehydrogenation of propane},}\ }\href
  {https://doi.org/10.1021/jacs.3c04030} {\bibfield  {journal} {\bibinfo
  {journal} {J. Am. Chem. Soc.}\ }\textbf {\bibinfo {volume} {145}},\ \bibinfo
  {pages} {14894--14902} (\bibinfo {year} {2023})}\BibitemShut {NoStop}%
\bibitem [{\citenamefont {Du}\ \emph {et~al.}(2023)\citenamefont {Du},
  \citenamefont {Damewood}, \citenamefont {Lunger}, \citenamefont {Millan},
  \citenamefont {Yildiz}, \citenamefont {Li},\ and\ \citenamefont
  {G{\'o}mez-Bombarelli}}]{du2023machine}%
  \BibitemOpen
  \bibfield  {author} {\bibinfo {author} {\bibfnamefont {X.}~\bibnamefont
  {Du}}, \bibinfo {author} {\bibfnamefont {J.~K.}\ \bibnamefont {Damewood}},
  \bibinfo {author} {\bibfnamefont {J.~R.}\ \bibnamefont {Lunger}}, \bibinfo
  {author} {\bibfnamefont {R.}~\bibnamefont {Millan}}, \bibinfo {author}
  {\bibfnamefont {B.}~\bibnamefont {Yildiz}}, \bibinfo {author} {\bibfnamefont
  {L.}~\bibnamefont {Li}},\ and\ \bibinfo {author} {\bibfnamefont
  {R.}~\bibnamefont {G{\'o}mez-Bombarelli}},\ }\bibfield  {title} {\enquote
  {\bibinfo {title} {Machine-learning-accelerated simulations to enable
  automatic surface reconstruction},}\ }\href
  {https://doi.org/10.5281/zenodo.7758174} {\bibfield  {journal} {\bibinfo
  {journal} {Nat. Comput. Sci.}\ }\textbf {\bibinfo {volume} {3}},\ \bibinfo
  {pages} {1034--1044} (\bibinfo {year} {2023})}\BibitemShut {NoStop}%
\bibitem [{\citenamefont {Kumari}, \citenamefont {Alexandrova},\ and\
  \citenamefont {Sautet}(2023)}]{kumari2023nature}%
  \BibitemOpen
  \bibfield  {author} {\bibinfo {author} {\bibfnamefont {S.}~\bibnamefont
  {Kumari}}, \bibinfo {author} {\bibfnamefont {A.~N.}\ \bibnamefont
  {Alexandrova}},\ and\ \bibinfo {author} {\bibfnamefont {P.}~\bibnamefont
  {Sautet}},\ }\bibfield  {title} {\enquote {\bibinfo {title} {Nature of
  zirconia on a copper inverse catalyst under co2 hydrogenation conditions},}\
  }\href@noop {} {\bibfield  {journal} {\bibinfo  {journal} {Journal of the
  American Chemical Society}\ }\textbf {\bibinfo {volume} {145}},\ \bibinfo
  {pages} {26350--26362} (\bibinfo {year} {2023})}\BibitemShut {NoStop}%
\bibitem [{\citenamefont {Zhang}, \citenamefont {Zandkarimi},\ and\
  \citenamefont {Alexandrova}(2020)}]{zhang2020ensembles}%
  \BibitemOpen
  \bibfield  {author} {\bibinfo {author} {\bibfnamefont {Z.}~\bibnamefont
  {Zhang}}, \bibinfo {author} {\bibfnamefont {B.}~\bibnamefont {Zandkarimi}},\
  and\ \bibinfo {author} {\bibfnamefont {A.~N.}\ \bibnamefont {Alexandrova}},\
  }\bibfield  {title} {\enquote {\bibinfo {title} {Ensembles of metastable
  states govern heterogeneous catalysis on dynamic interfaces},}\ }\href
  {https://doi.org/10.1021/acs.accounts.9b00531} {\bibfield  {journal}
  {\bibinfo  {journal} {Acc. Chem. Res.}\ }\textbf {\bibinfo {volume} {53}},\
  \bibinfo {pages} {447--458} (\bibinfo {year} {2020})}\BibitemShut {NoStop}%
\bibitem [{\citenamefont {Karlberg}\ \emph {et~al.}(2007)\citenamefont
  {Karlberg}, \citenamefont {Jaramillo}, \citenamefont {Skulason},
  \citenamefont {Rossmeisl}, \citenamefont {Bligaard},\ and\ \citenamefont
  {N{\o}rskov}}]{karlberg2007cyclic}%
  \BibitemOpen
  \bibfield  {author} {\bibinfo {author} {\bibfnamefont {G.}~\bibnamefont
  {Karlberg}}, \bibinfo {author} {\bibfnamefont {T.}~\bibnamefont {Jaramillo}},
  \bibinfo {author} {\bibfnamefont {E.}~\bibnamefont {Skulason}}, \bibinfo
  {author} {\bibfnamefont {J.}~\bibnamefont {Rossmeisl}}, \bibinfo {author}
  {\bibfnamefont {T.}~\bibnamefont {Bligaard}},\ and\ \bibinfo {author}
  {\bibfnamefont {J.~K.}\ \bibnamefont {N{\o}rskov}},\ }\bibfield  {title}
  {\enquote {\bibinfo {title} {Cyclic voltammograms for h on pt (111) and pt
  (100) from first principles},}\ }\href
  {https://doi.org/10.1103/PhysRevLett.99.126101} {\bibfield  {journal}
  {\bibinfo  {journal} {Phys. Rev. Lett.}\ }\textbf {\bibinfo {volume} {99}},\
  \bibinfo {pages} {126101} (\bibinfo {year} {2007})}\BibitemShut {NoStop}%
\bibitem [{\citenamefont {Zhou}\ \emph {et~al.}(2022)\citenamefont {Zhou},
  \citenamefont {Zhu}, \citenamefont {Scheffler},\ and\ \citenamefont
  {Ghiringhelli}}]{zhou2022ab}%
  \BibitemOpen
  \bibfield  {author} {\bibinfo {author} {\bibfnamefont {Y.}~\bibnamefont
  {Zhou}}, \bibinfo {author} {\bibfnamefont {C.}~\bibnamefont {Zhu}}, \bibinfo
  {author} {\bibfnamefont {M.}~\bibnamefont {Scheffler}},\ and\ \bibinfo
  {author} {\bibfnamefont {L.~M.}\ \bibnamefont {Ghiringhelli}},\ }\bibfield
  {title} {\enquote {\bibinfo {title} {Ab initio approach for thermodynamic
  surface phases with full consideration of anharmonic effects: The example of
  hydrogen at si (100)},}\ }\href
  {https://doi.org/10.1103/PhysRevLett.128.246101} {\bibfield  {journal}
  {\bibinfo  {journal} {Phys. Rev. Lett.}\ }\textbf {\bibinfo {volume} {128}},\
  \bibinfo {pages} {246101} (\bibinfo {year} {2022})}\BibitemShut {NoStop}%
\bibitem [{\citenamefont {Zhou}, \citenamefont {Scheffler},\ and\ \citenamefont
  {Ghiringhelli}(2019)}]{zhou2019determining}%
  \BibitemOpen
  \bibfield  {author} {\bibinfo {author} {\bibfnamefont {Y.}~\bibnamefont
  {Zhou}}, \bibinfo {author} {\bibfnamefont {M.}~\bibnamefont {Scheffler}},\
  and\ \bibinfo {author} {\bibfnamefont {L.~M.}\ \bibnamefont {Ghiringhelli}},\
  }\bibfield  {title} {\enquote {\bibinfo {title} {Determining surface phase
  diagrams including anharmonic effects},}\ }\href
  {https://doi.org/10.1103/PhysRevB.100.174106} {\bibfield  {journal} {\bibinfo
   {journal} {Phys. Rev. B}\ }\textbf {\bibinfo {volume} {100}},\ \bibinfo
  {pages} {174106} (\bibinfo {year} {2019})}\BibitemShut {NoStop}%
\bibitem [{\citenamefont {Masel}(1996)}]{masel_1996}%
  \BibitemOpen
  \bibfield  {author} {\bibinfo {author} {\bibfnamefont {R.}~\bibnamefont
  {Masel}},\ }\href@noop {} {\emph {\bibinfo {title} {Principles of adsorption
  and reaction on solid surfaces}}}\ (\bibinfo  {publisher} {Wiley and Sons},\
  \bibinfo {address} {Hoboken},\ \bibinfo {year} {1996})\BibitemShut {NoStop}%
\bibitem [{\citenamefont {Zhang}, \citenamefont {Blum},\ and\ \citenamefont
  {Reuter}(2007)}]{zhang_2007}%
  \BibitemOpen
  \bibfield  {author} {\bibinfo {author} {\bibfnamefont {Y.}~\bibnamefont
  {Zhang}}, \bibinfo {author} {\bibfnamefont {V.}~\bibnamefont {Blum}},\ and\
  \bibinfo {author} {\bibfnamefont {K.}~\bibnamefont {Reuter}},\ }\bibfield
  {title} {\enquote {\bibinfo {title} {Accuracy of ﬁrst-principles lateral
  interactions: Oxygen at pd(100)},}\ }\href@noop {} {\bibfield  {journal}
  {\bibinfo  {journal} {Phys. Rev. B}\ }\textbf {\bibinfo {volume} {75}},\
  \bibinfo {pages} {235406} (\bibinfo {year} {2007})}\BibitemShut {NoStop}%
\bibitem [{\citenamefont {{\AA}ngqvist}\ \emph {et~al.}(2019)\citenamefont
  {{\AA}ngqvist}, \citenamefont {Mu{\~n}oz}, \citenamefont {Rahm},
  \citenamefont {Fransson}, \citenamefont {Durniak}, \citenamefont {Rozyczko},
  \citenamefont {Rod},\ and\ \citenamefont {Erhart}}]{aangqvist2019icet}%
  \BibitemOpen
  \bibfield  {author} {\bibinfo {author} {\bibfnamefont {M.}~\bibnamefont
  {{\AA}ngqvist}}, \bibinfo {author} {\bibfnamefont {W.~A.}\ \bibnamefont
  {Mu{\~n}oz}}, \bibinfo {author} {\bibfnamefont {J.~M.}\ \bibnamefont {Rahm}},
  \bibinfo {author} {\bibfnamefont {E.}~\bibnamefont {Fransson}}, \bibinfo
  {author} {\bibfnamefont {C.}~\bibnamefont {Durniak}}, \bibinfo {author}
  {\bibfnamefont {P.}~\bibnamefont {Rozyczko}}, \bibinfo {author}
  {\bibfnamefont {T.~H.}\ \bibnamefont {Rod}},\ and\ \bibinfo {author}
  {\bibfnamefont {P.}~\bibnamefont {Erhart}},\ }\bibfield  {title} {\enquote
  {\bibinfo {title} {Icet--a python library for constructing and sampling alloy
  cluster expansions},}\ }\href {https://doi.org/10.1002/adts.201900015}
  {\bibfield  {journal} {\bibinfo  {journal} {Adv. Theory Simul.}\ }\textbf
  {\bibinfo {volume} {2}},\ \bibinfo {pages} {1900015} (\bibinfo {year}
  {2019})}\BibitemShut {NoStop}%
\end{thebibliography}%
\end{document}